\documentclass[letterpaper, 10 pt, conference]{ieeeconf}  

\IEEEoverridecommandlockouts                              
                                                          
\usepackage{amsmath}
\usepackage{amssymb}
\usepackage{amscd}
\usepackage{amsfonts}
\usepackage{dsfont}
\usepackage{graphicx}%
\usepackage{subcaption}
\usepackage{fancyhdr}

\usepackage{cite}
\usepackage{color}
\usepackage[ruled]{algorithm2e}
\usepackage{ifthen}
\usepackage{algorithmic}
\newtheorem{lemma}{Lemma}
\newtheorem{theorem}{Theorem}

\newtheorem{definition}{Definition}

\newtheorem{corollary}{Corollary}
\newtheorem{remark}{Remark}

\usepackage{cite}
\usepackage{color}
\usepackage[ruled]{algorithm2e}
\usepackage{ifthen}
\usepackage{algorithmic}

\def\scr#1{{\cal #1}} 
\ifodd 1
\def\eq#1{\begin{equation}#1\end{equation}}
\newcommand{\R}{{\rm I\!R}}
\newcommand{\bbb}{\mathbb}

\else

\newcommand{\com}[1]{}
\newcommand{\clar}[1]{}
\newcommand{\response}[1]{}
\fi

\def\rep#1{(\ref{#1})}

\usepackage{hyperref}

\def\qed{ \rule{.08in}{.08in}}

\title{Resilient Constrained Consensus over Complete Graphs\\ via Feasibility Redundancy
}

\author{Jingxuan Zhu  \hspace{.3in} Yixuan Lin \hspace{.3in} Alvaro Velasquez \hspace{.3in} Ji Liu
\thanks{The research of J. Zhu, Y. Lin and J. Liu was supported in part
by the US Army Research Laboratory (ARL) Cooperative
Agreement W911NF-21-2-0098.
J.~Zhu and Y.~Lin are with the Department of Applied Mathematics and Statistics at Stony Brook University (\texttt{\{jingxuan.zhu,yixuan.lin.1\}@stonybrook.edu}).
J.~Liu is with the Department of Electrical and Computer Engineering at Stony Brook University (\texttt{ji.liu@stonybrook.edu}).
A.~Velasquez is with the Information Directorate of the Air Force Research Laboratory (\texttt{alvaro.velasquez.1@us.af.mil}).
}
}

\begin{document}

\maketitle
\thispagestyle{empty}
\pagestyle{empty}

\begin{abstract}
This paper considers a resilient high-dimensional constrained consensus problem and studies a resilient distributed algorithm for complete graphs. For convex constrained sets with a singleton intersection, a sufficient condition on feasibility redundancy and set regularity for reaching a desired consensus exponentially fast in the presence of Byzantine agents is derived, which can be directly applied to polyhedral sets. A necessary condition on feasibility redundancy for the resilient constrained consensus problem to be solvable is also provided. 
\end{abstract}


\section{Introduction}

There have been considerable advances in distributed control and computation algorithms since the study of the flocking problem \cite{flocking}, which is also known as the consensus problem \cite{murrarysurvey}.  
Almost all the distributed algorithms include consensus 
as a key component for information fusion. 
Therefore, devising resilient consensus algorithms is the first step toward resilient distributed algorithms in the presence of adversarial agents.
The research on consensus in the presence of faulty processes can be dated back to \cite{FiLyPa85,DoLyPiStWe86} which established fundamental limits and impossibility results for complete communication graphs.
Recently, there has been considerable interest in designing resilient consensus algorithms for general, incomplete communication graphs. 
Notable examples are \cite{VaTsLi12,LeZhKoSu13}
which solve the scalar-valued resilient consensus problem using graph~redundancy. 

In many distributed algorithms such as distributed optimization and machine learning, agents share relevant local information by exchanging real-valued vectors that are frequently high-dimensional. 
In high-dimensional settings, given a set of both non-faulty and faulty vectors (e.g., caused by Byzantine attacks), a fundamental problem, the ``Byzantine vector consensus'' problem \cite{Va13,Va14}, is to find a vector which lies in the convex hull of the non-faulty vectors.
Necessary and sufficient conditions have been established in \cite{Va13} to guarantee the existence of the desired vector based on Tverberg's theorem \cite{tverberg,tverbergpoint}, and corresponding algorithms have been developed in \cite{Va14} for general graphs.
Such algorithms require a number of redundant agents that increase linearly in not only the number of adversarial agents but also the dimension of the consensus vector \cite{Va13,Va14}. 
One recent approach appeals to the idea of centerpoints \cite{center} which improves the
necessary and sufficient conditions for resilient vector consensus.
However, computing the desired intersection of convex hulls, Tverberg partition or centerpoint in high dimensions is computationally expensive. 
In a word, designing a resilient vector consensus scheme applicable to high-dimensional distributed algorithms in multi-agent networks is challenging, and there has been vast literature on this area over the past decade, which cannot be thoroughly reviewed here.
Among all the existing literature, we are motivated by a recent paper \cite{idea} which studies a resilient distributed convex optimization problem over complete graphs; the paper proposes a clever and computationally efficient way for each agent to ``filter'' its received values and solves the resilient optimization problem under the assumptions of cost function redundancy (see Definition 2 in \cite{idea}) and strong convexity of cost functions, the latter implying a unique optimal solution. 

It turns out that the algorithm proposed in \cite{idea} makes use of the idea of {\em constrained consensus} \cite{constrain,cdc14constrain,tacrate}, whose goal is for a group of agents to reach a consensus while each agent's state is constrained to be in its local convex set. Such constraints are inevitable and important in a number of applications including multi-robot motion planning and alignment problems, where the constraints represent environmental limitations or ``safety'' requirements \cite{planning}.
To the best of our knowledge, resilient constrained consensus has never been studied, and this paper appears to be
the first study of resilient discrete-time constrained consensus. 
Since the problem is challenging, we focus on complete graphs as a promising starting point of this line of research.

\vspace{.1in}

\section{Problem and Results}

Consider a network of $n$ agents including up to $f$ {\em Byzantine} agents which are able to transmit arbitrary values to others and capable of sending conflicting values to different neighbors at any time. 
The communication relations among the $n$ agents is assumed to be an $n$-vertex complete graph $\bbb G$; in other words, each agent can communicate with any other agent. Let $\scr N$ be the set of all $n$ agents, $\scr H$ be the set of normal agents with $|\scr H|\ge n-f$, and $\scr F$ be the set of Byzantine agents with $|\scr F|\le f$. 
The number $f$ is known across the network, yet exactly which agents are Byzantine is unknown to any normal agent.
The state of each normal agent~$i\in\scr H$ is constrained to be in a closed convex set $\scr X_i\in\R^m$ and $\scr X\triangleq\bigcap_{i\in\scr H}\scr X_i$ is nonempty. The goal of the resilient constrained consensus problem is to devise an algorithm for each normal agent which guarantees that all the normal agents will reach a consensus at some point in~$\scr X$.

We consider the following discrete-time synchronous algorithm for the problem described above. 
Let $x_i(t)\in\R^m$ denote agent $i$'s value at time $t$. Since $\bbb G$ is complete, each agent receives values from all other agents at each time. Let $x_{ji}(t)$ denote the value received by agent $i$ from agent $j$ at time $t$.\footnote{If agent $j$ is Byzantine, it is possible that $x_{ji}(t)\neq x_{jk}(t)$ for any time $t$ and $i\neq k$.} 
At each time $t$, each normal agent $i\in\scr H$ first filters its received values as follows:
Agent $i$ computes the Euclidean distance between its own current value $x_i(t)$ and each of received values $x_{ji}(t)$, $j\in\scr N\setminus \{i\}$, and removes the furthest $f$ values, with ties broken arbitrarily. Let $\scr M_i(t)$ denote the set of agents in $\scr N\setminus \{i\}$ whose values are retained by agent $i$ at time $t$. 
Then, agent $i$ updates its state by setting 
\begin{align}
    x_i(t+1)=\mathrm P_{\scr X_i}\bigg[x_i(t) + \alpha\sum_{j\in\scr M_i(t)}\left(x_{ji}(t)-x_i(t)\right)\bigg],\label{update}
\end{align}
where $\alpha$ is a positive stepsize, and $\mathrm P_{\scr C}[\cdot]$ denotes the Euclidean projection on a closed convex set $\scr C$, i.e., $\mathrm P_{\scr C}[x] = \arg\min_{y \in \scr C} \| x- y \|$, where $\|\cdot\|$ denotes the Euclidean norm. Without Byzantine agents and projection operations, algorithm \eqref{update} simplifies to $x_i(t+1)=x_i(t) + \alpha\sum_{j\in\scr N_i}(x_{j}(t)-x_i(t))$, a conventional (unconstrained) consensus algorithm \cite{murrarysurvey}, where $\scr{N}_i$ denotes the set of neighbors of agent $i$. Without Byzantine agents, algorithm \eqref{update} becomes $x_i(t+1)=\mathrm P_{\scr X_i}[x_i(t) + \alpha\sum_{j\in\scr N_i}(x_{j}(t)-x_i(t))]$, a special case of the discrete-time constrained consensus algorithm \cite{constrain,tacrate}. This paper aims to investigate when algorithm~\eqref{update} will guarantee a consensus among all normal~agents. 

\vspace{.05in}

\begin{remark}
Algorithm~\eqref{update} was studied in \cite{idea} for solving a resilient distributed convex optimization problem in which $\scr X_i$ is the optimal solution set of agent $i$'s local cost function; thus the conditions provided there rely on Lipschitz smoothness and strong convexity of the cost functions, which do not exist in constrained consensus. With these in mind, we study algorithm \eqref{update} from a set property perspective. 
Thus, the results obtained in \cite{idea} cannot be applied to here, and vice versa. 
\hfill$\Box$
\end{remark}

\vspace{.05in}

To make the resilient constrained consensus problem solvable, certain feasibility redundancy is necessary.

\vspace{.05in}

\begin{definition}\label{red}
The set of normal agents, $\scr H$, 
is called {\em $k$-redundant} if for any subset $\scr S\subseteq \scr H$ such that $|\scr S|\ge n-k$, there holds 
$
    \bigcap_{i\in \scr S}\scr X_i=\bigcap_{i\in\scr H}\scr X_i
$.
\end{definition}

\vspace{.05in}

The definition is prompted by Definition 2 in \cite{reviewer}.

\vspace{.05in}

\begin{theorem} \label{thm:2f_redundancy}
Suppose that there are $f$ Byzantine agents. Then, all the normal agents can reach a consensus at some point in $\scr X$ via algorithm \eqref{update} only if the set of normal agents $\scr H$ is $2f$-redundant.
\end{theorem}

\vspace{.05in}

The above theorem provides a necessary condition for algorithm \eqref{update} to solve the resilient constrained consensus problem. In the sequel, we will present a sufficient condition for solving the problem. 


Let ${\rm dist}(x, \scr C)$ denote the distance of a point $x$ to a closed convex set $\scr C$, i.e., ${\rm dist}(x, \scr C) = \| x - \mathrm{P}_{\scr C} [x]\|$.


\begin{theorem} \label{thm:single}
Suppose that $\scr X = \{x^*\}$ is a singleton and there are at most $f$ Byzantine agents.
Suppose that the set of normal agents $\scr H$ is $k$-redundant and
there exists a positive constant $\mu$ such that for any $x\in\R^m$ and $\scr S\subseteq \scr H$ with $|\scr S|\ge n-k$,
\begin{equation}\label{mu}
    \max_{i\in\scr S}\{{\rm dist}(x,\scr X_i)\}\ge \mu \;{\rm dist}\Big(x,\;\bigcap_{i\in \scr S}\scr X_i\Big).
\end{equation}
If $k>\frac{4f}{\mu^2}+2f-1$ and
$\alpha < (\mu^2 k - 2f\mu^2 - 4f + \mu^2)/(4|\scr H|^3)$,
then there exists a constant $\rho\in(0,1)$ for which 
\begin{align*}
    \sum_{i\in\scr H}\|x_i(t)-x^*\|^2\le \rho^t\sum_{i\in\scr H}\|x_i(0)-x^*\|^2, \;\;\; t\ge 0.
\end{align*}
\end{theorem}

\vspace{.1in}

The theorem implies that all the normal agents will reach a consensus at $x^*$ exponentially fast.


From Definition~\ref{red}, the $k$-redundancy requirement can be met only if the total number of agents $n\ge k+f$; in other words, the number of normal agents cannot be smaller than $k$.
It is easy to see that the positive constant $\mu$ in \rep{mu} must satisfy $\mu\le 1$. More discussion on the role of $\mu$ in constrained consensus can be found in \cite{tacrate}. 
It is also easy to show that the upper bound of $\alpha$ in Theorem~\ref{thm:single} is strictly positive. 


Although condition \rep{mu} in Theorem~\ref{thm:single} looks restrictive at a glance, it immediately satisfies if each $\scr X_i$ is a polyhedral set, i.e., $\scr X_i = \{ x \;|\; a_i^\top x \le b_i \}$ with $a_i\in\R^m$ and $b_i\in\R$, as proved in \cite{hoffman}. We thus have the following result. 

\vspace{.05in}

\begin{corollary}
Suppose that $\scr X_i = \{ x \;|\; a_i^\top x \le b_i \}$, where $a_i\in\R^m$ and $b_i\in\R$ for all $i\in\scr H$, and $\bigcap_{i\in\scr H}\scr X_i = \{x^*\}$ is a singleton. 
Suppose that there are at most $f$ Byzantine agents and
the set of normal agents $\scr H$ is $k$-redundant. Then, there exists a positive constant $\mu$ such that \rep{mu} holds for any $x\in\R^m$ and $\scr S\subseteq \scr H$ with $|\scr S|\ge n-k$.
If, furthermore, $k>\frac{4f}{\mu^2}+2f-1$ and
$\alpha < (\mu^2 k - 2f\mu^2 - 4f + \mu^2)/(4|\scr H|^3)$,
there exists a constant $\rho\in(0,1)$ for which 
\begin{align*}
    \sum_{i\in\scr H}\|x_i(t)-x^*\|^2\le \rho^t\sum_{i\in\scr H}\|x_i(0)-x^*\|^2, \;\;\; t\ge 0.
\end{align*}
\end{corollary}

\vspace{.1in}

\section{Analysis}

In this section, we provide the proofs of Theorems \ref{thm:2f_redundancy} and~\ref{thm:single}.

\subsection{Proof of Theorem~\ref{thm:2f_redundancy}}
We prove the theorem by contradiction. Suppose that $\scr H$ does not satisfy $2f$-redundancy, then we only need to show it is possible for the normal agents to converge to a consensus point outside of $\scr X$ or they do not reach consensus.

Since $\scr H$ does not satisfy $2f$-redundancy, there exists $\hat{\scr T} \subset \scr H,$ such that $|\hat{\scr T}|\ge n-2f$ and  $\bigcap_{i \in \hat{\scr T}}\scr X_i \supset \scr X,$ i.e., $\exists\;x\in\bigcap_{i \in \hat{\scr T}}\scr X_i\setminus\scr X.$ Then we can find a subset $\scr T \subseteq \hat{\scr T},$ such that $|{\scr T}| = n-2f,$ and we have  $x\in\bigcap_{i \in {\scr T}}\scr X_i\setminus\scr X$. The existence of $\hat{\scr T}$ also implies that there exists at least a $j\in\scr H\setminus\hat{\scr T},$ such that $x\notin \scr X_j.$
Now consider the case when there are exactly $f$ Byzantine agents, i.e., $|\scr F| = f,$ and that $x_i(0)=x$ for all $i\in \scr T$ and $x_{jh}(t)=x$ for all $j\in\scr F, h\in\scr H$ and $t\ge0.$ We will use induction to show $x_i(t) = x$ for all $i \in \scr T$ and $t > 0.$ 

From the definition of $\scr M_i(t),$ it is easy to show that for all $i \in \scr T$ and $j \in \scr M_i(0)$, we have $x_{ji}(0) = x $ and $x_i(1) = \mathrm P_{\scr X_i}[x_i(0) + \alpha\sum_{j\in\scr M_i(0)}\left(x_{ji}(0)-x_i(0)\right)] = x$.

Suppose $x_i(t) = x$ for all $i \in \scr T,$ next we will validate that $x_i(t + 1) = x$ for all $i \in \scr T.$  Since for $i\in\scr T,$ we have $|\scr T\setminus\{i\}| + |\scr F| = n - f - 1 = |\scr M_i(t)|,$ and since for any $h\in\scr T\setminus\{i\}\bigcup\scr F,$ it holds that $\|x_i(t) - x_{hi}(t)\| = 0,$ we imply from the filtering policy that $\scr M_i(t) = \scr T\setminus\{i\}\bigcup\scr F,$ then from \eqref{update}, $x_i(t+1) = x,$ which completes the induction.

The above analysis implies that for all $i\in\scr T,$ $x_i(t)$ converges to $x,$ along with the fact that $x\notin \scr X_j,$ we conclude that normal agents cannot reach a consensus in the proposed example,
which completes the proof.
\hfill $\qed$

\subsection{Proof of Theorem~\ref{thm:single}}
To prove Theorem~\ref{thm:single}, we need the following lemmas.

\vspace{.05in}

\begin{lemma} \label{lemma:single_main_eq}
Suppose that the set of normal agents $\scr H$ is $k$-redundant and
there exists a positive constant $\mu$ such that \eqref{mu} holds for any $x\in\R^m$ and $\scr S\subseteq \scr H$ with $|\scr S|\ge n-k$. Let $x_j$ be an arbitrary element in $\scr X_j,\;\forall j\in\scr H,$ and $\scr S$ be any set that contains at least $n-k$ normal agents. Then $\exists\; 0<\mu\le1$ such that for $\forall i\in\scr H,$ 
\begin{align} \label{eq:largestdistance}
    \max_{j\in\scr S}\|x_i-x_j\|^2\ge \mu^2 \|x_i-x^*\|^2.
\end{align}
\end{lemma}

\vspace{.1in}

{\em Proof of Lemma~\ref{lemma:single_main_eq}:}
Let $\scr S$ be an arbitrary subset of $\scr H$ that satisfies $|\scr S|\ge n-k.$ Since $\scr H$ is $k$-redundant,  $\bigcap_{j\in \scr S}\scr X_j=\bigcap_{j\in\scr H}\scr X_j=\scr X =\{x^*\},$
then using \eqref{mu}, we have
\begin{align} \label{eq:single_lemma_1}
    \max_{j\in\scr S}\{{\rm dist}(x_i,\scr X_j)\}\ge {\mu}\;{\rm dist}(x,\scr X)= {\mu} \|x_i-x^*\|.
\end{align}
It is easy to see that for any $x_j\in\scr X_j, j\in\scr H$ we have
\begin{align}  \label{eq:single_lemma_2}
    \|x_i-x_j\|\ge {\rm dist}(x_i,X_j).
\end{align}
Combining \eqref{eq:single_lemma_1} and \eqref{eq:single_lemma_2}, we obtain that
\begin{align*}
     \max_{j\in\scr S}\|x_i-x_j\|^2\ge  \max_{j\in\scr S}\{{\rm dist}^2(x_i,\scr X_j)\}\ge \mu^2 \|x_i-x^*\|^2.
\end{align*}
This completes the proof.
\hfill$\qed$

\vspace{.1in}


To proceed, let $V(t) = \|\mathrm{P}_{\scr X_i}[x_i(t)] - x^*\|^2.$ From algorithm \eqref{update}, we obtain that
$$
    V_i(t+1)
    =\Big\|\mathrm{P}_{\scr X_i}\big[x_i(t)+ \alpha \sum_{j\in\scr M_i(t)}(x_{ji}(t) - x_i(t))\big] - x^*\Big\|^2.
$$
Let $\langle x, y \rangle = x^\top y $ for any $x, y \in \R^m$. Due to non-expansion property of Euclidean projection onto a convex set, 
\begin{align*}
    V_i(t+1)
    & \le \Big\| x_i(t) - x^* + \alpha \sum_{j\in\scr M_i(t)}(x_{ji}(t) - x_i(t)) \Big\|^2 \\
    & = V_i(t) + \alpha^2\Big\|\sum_{j\in\scr M_i(t)}(x_{ji}(t) - x_i(t)) \Big\|^2 \nonumber\\ 
    &\;\;\;\;- 2 \alpha \Big \langle x_i(t)-x^* , \sum_{j\in\scr M_i(t)}(x_i(t)-x_{ji}(t))\Big\rangle .
\end{align*}
Let 
\eq{
    \phi_i(t)=\Big \langle x_i(t)-x^* , \sum_{j\in\scr M_i(t)}(x_i(t)-x_{ji}(t))\Big\rangle,\label{eq:phi}
}
and 
\eq{
    \psi_i(t)=\Big\|\sum_{j\in\scr M_i(t)}(x_{ji}(t) - x_i(t)) \Big\|^2.\label{eq:psi}
}
Combining these together, we have
\begin{align}
    V_i(t+1)\le V_i(t) - 2 \alpha \phi_i(t) + \alpha^2 \psi_i(t). \label{eq:Vi_iteration}
\end{align}
Summing up \eqref{eq:Vi_iteration} for all $i\in\scr H,$ we obtain that
\begin{align}
    V(t+1)\le V(t) - 2\alpha \sum_{i\in\scr H} \phi_i(t) + \alpha^2 \sum_{i\in\scr H} \psi_i(t).\label{eq:V:iteration}
\end{align}
For all $i\in\scr H,$ let $\scr H_i=\scr H\setminus\{i\},$  and $\scr L_i(t)$ be the set of normal agents in the filtered set $\scr M_i(t),$ i.e., $\scr L_i(t)=\scr M_i(t)\bigcap\scr H_i$, and 
$\scr F_i(t)=\scr M_i(t)\setminus\scr L_i(t).$ Note that $|\scr H|-f-1\le|\scr L_i(t)|\le n-f-1.$ 

\vspace{.05in}

\begin{lemma} \label{lemma:lower_bound_S}
For any agent $i \in \scr H$, let 
\begin{align}
    e_i(t)&=\sum_{j\in\scr F_i(t)}(x_i(t)-x_{ji}(t))-\sum_{j\in \scr H_i\setminus\scr L_i(t)}(x_i(t)-x_{ji}(t))\label{e}
\end{align}    
and
\begin{align}
    S_i(t)&= \frac{1}{2} \sum_{j\in\scr H} \| x_i(t) - x_j(t) \|^2 + \langle x_i(t) - x^* , e_i(t) \rangle. \label{Si}
\end{align}
Then, 
\begin{align}
    S_i(t)
    &\ge \frac{1}{2}\sum_{j\in\scr L_i(t)}\|x_i(t)-x_j(t)\|^2 \nonumber\\ &\;\;\;\;
    -2\sum_{j\in \scr H_i\setminus\scr L_i(t)}\|x_i(t)-x^*\|^2,\label{eq:Si_upp_1}
\end{align}
and
\begin{align}
    \sum_{i\in\scr H}\phi_i(t)&=\sum_{i\in\scr H}S_i(t).\label{phi&Si}
\end{align}
\end{lemma}

\vspace{.15in}

{\em Proof of Lemma~\ref{lemma:lower_bound_S}:}
From the definition of $\scr L_i(t)$ and $\scr F_i(t)$,
\begin{align}
    &\;\;\;\;\sum_{j\in\scr M_i(t)} (x_i(t)-x_{ji}(t)) \nonumber\\ 
    &=\sum_{j\in\scr L_i(t)}(x_i(t)-x_{ji}(t))+\sum_{j\in\scr F_i(t)}(x_i(t)-x_{ji}(t))\label{eq:single_sum_split}\\
    &=\;\sum_{j\in\scr H_i}(x_i(t)-x_j(t))-\sum_{j\in \scr H_i\setminus\scr L_i(t)}(x_i(t)-x_{ji}(t))\nonumber\\ 
    &\;\;\;\;\;\;+\sum_{j\in\scr F_i(t)}(x_i(t)-x_{ji}(t)) \nonumber.
\end{align}
From the definition of $e_i(t)$ given in \eqref{e}, we have
$
    \sum_{j\in\scr M_i(t)}(x_i(t)-x_{ji}(t))=\sum_{j\in\scr H_i}(x_i(t)-x_j(t))+ e_i(t).
$
Substituting it in \eqref{eq:phi}, we obtain that
\begin{align*}
    \phi_i(t)&=\Big\langle x_i(t)-x^*,\sum_{j\in\scr H_i}(x_i(t)-x_j(t))\Big\rangle \\
    &\;\;\;\;\; +\langle x_i(t)-x^*,e_i(t)\rangle.
\end{align*}
Therefore, 
\begin{align*}
    \sum_{i\in\scr H}\phi_i(t)
    &=\sum_{i\in\scr H}\Big(\big\langle x_i(t)-x^*,
    \sum_{j\in\scr H_i}(x_i(t)-x_j(t))\big\rangle \\
    &\qquad\quad+\langle x_i(t)-x^*,e_i(t)\rangle\Big).
\end{align*}
Note that as $x_i(t)-x_j(t)=x_i(t)-x^*-(x_j(t)-x^*)$ and that graph $\bbb G$ is complete,
\begin{align}
    &\;\;\;\; \sum_{i\in\scr H}\Big\langle x_i(t)-x^*,\sum_{j\in\scr H_i}(x_i(t)-x_j(t))\Big\rangle \nonumber\\
    &=\sum_{i\in\scr H}\sum_{j\in\scr H}\langle x_i(t)-x^*,x_i(t)-x_j(t)\rangle\label{eq1}\\
    &=\sum_{i\in\scr H}\sum_{j\in\scr H}\|x_i(t)-x_j(t)\|^2 \nonumber\\
    &\;\;\;\; +\sum_{i\in\scr H}\sum_{j\in\scr H}\langle x_j(t)-x^*,x_i(t)-x_j(t)\rangle\nonumber\\
    &=\sum_{i\in\scr H}\sum_{j\in\scr H}\|x_i(t)-x_j(t)\|^2\nonumber\\ 
    &\;\;\;\; -\sum_{i\in\scr H}\sum_{j\in\scr H}\langle x_i(t)-x^*,x_i(t)-x_j(t)\rangle, \label{eq2}
\end{align}
where we substitute $i$ and $j$ in the last step.
Combining \eqref{eq1} and \eqref{eq2}, we have
\begin{align*}
    &\;\;\;\;\;\sum_{i\in\scr H}\Big\langle x_i(t)-x^*,\sum_{j\in\scr H_i}(x_i(t)-x_j(t))\Big\rangle \nonumber\\ 
     &=\;\frac{1}{2}\sum_{i\in\scr H}\sum_{j\in\scr H}\|x_i(t)-x_j(t)\|^2.
\end{align*}
From the definition of $S_i(t)$ in \eqref{Si}, we have $\sum_{i\in\scr H}\phi_i(t)=\sum_{i\in\scr H}S_i(t).$
From \eqref{e}, we obtain that $\|e_i(t)\|\le \sum_{j\in\scr F_i(t)}\|x_i(t)-x_{ji}(t)\|+\sum_{j\in \scr H_i\setminus\scr L_i(t)}\|x_i(t)-x_j(t)\|.$
Recall from the algorithm, it holds that $ \|x_i(t)-x_{ki}(t)\|\le \|x_i(t)-x_j(t)\|$ for $k\in\scr F_i(t)$ and $j\in\scr H_i\setminus\scr L_i(t).$
Since $|\scr F_i(t)|=|\scr H_i\setminus \scr L_i(t)|,$ we have
\begin{align}
    \sum_{j\in\scr F_i(t)} \|x_i(t)-x_{ji}(t)\|\le \sum_{j\in\scr H_i\setminus \scr L_i(t)}\|x_i(t)-x_j(t)\|.\label{Bequality}
\end{align}
Thus, $\|e_i(t)\|\le 2\sum_{j\in \scr H_i\setminus\scr L_i(t)}\|x_i(t)-x_j(t)\| .$
Moreover, using the Cauchy-Schwartz inequality, we have 
\begin{align*}
    &\langle x_i(t)-x^*,e_i(t)\rangle  \;\ge\; -\|x_i(t)-x^*\|\cdot\|e_i(t)\| \\
    &\ge\;-2\sum_{j\in \scr H_i\setminus\scr L_i(t)}\|x_i(t)-x^*\|\cdot\|x_i(t)-x_j(t)\|\\
     &\ge\; -\sum_{j\in \scr H_i\setminus\scr L_i(t)}\Big(2\|x_i(t)-x^*\|^2+\frac{1}{2}\|x_i(t)-x_j(t)\|^2\Big).
\end{align*}
Substituting from above in \eqref{Si}, we obtain that
\begin{align*}
    &\;\;\;\;\;S_i(t) \\
    &\ge \frac{1}{2}\sum_{j\in\scr H}\|x_i(t)-x_j(t)\|^2-\frac{1}{2}\sum_{j\in \scr H_i\setminus\scr L_i(t)}\|x_i(t)-x_j(t)\|^2\\
    &\;\;\;\;-2\sum_{j\in \scr H_i\setminus\scr L_i(t)}\|x_i(t)-x^*\|^2\\
    &=\frac{1}{2}\sum_{j\in\scr L_i(t)}\|x_i(t)-x_j(t)\|^2-2\sum_{j\in \scr H_i\setminus\scr L_i(t)}\|x_i(t)-x^*\|^2.
\end{align*}
This completes the proof.
\hfill$\qed$

\vspace{.05in}

\begin{lemma} \label{lemma:single_pt_set_phi}
Suppose that the set of normal agents $\scr H$ is $k$-redundant and there exists a positive constant $\mu$ such that \eqref{mu} holds for any $x\in\R^m$ and $\scr S\subseteq \scr H$ with $|\scr S|\ge n-k$.
Suppose that $\bigcap_{i\in\scr H}\scr X_i = \{x^*\}$. If $k>\frac{4f}{\mu^2}+(2f-1)$, then
\begin{align} \label{eq:lower_bound_sum_phi}
    \sum_{i\in\scr H}\phi_i(t)\ge\Big(\frac{\mu^2}{2}k-\frac{4f+2f\mu^2-\mu^2}{2}\Big)V(t).
\end{align}
\end{lemma}

\vspace{.1in}

{\em Proof of Lemma~\ref{lemma:single_pt_set_phi}:}
Since $|\scr L_i(t)|\ge n-2f-1$, $f \ge 1$ and $ n - k \le  n - \frac{4f}{\mu^2} - ( 2f - 1 ) \le n - 6f + 1   $, we have $ n - k \le  |\scr L_i(t) \cup \{ i \} |   $. Thus, there exists a subset $\hat{\scr L}_i(t) \subset \scr L_i(t)\cup \{ i \}$ with $|\hat{\scr L}_i(t)| = n-k$ such that for any agent $j \in \scr L_i(t)\cup \{ i \} \setminus \hat{\scr L}_i(t) $, we have
$
    \max_{k \in \hat{\scr L}_i(t)} \| x_i - x_k \|^2 \le \| x_i - x_j \|^2.
$
Let $\tilde{\scr L}_i(t) = \scr L_i(t)\cup \{ i \} \setminus \hat{\scr L}_i(t)$ for all $i$ and $t\ge0$.
By using \eqref{eq:largestdistance} in Lemma~\ref{lemma:single_main_eq}, we have
$
    \max_{j\in\hat{\scr L}_i(t)}\|x_i(t)-x_j(t)\|^2
    \ge \mu^2 \|x_i(t)-x^*\|^2,
$
and
\begin{align*}
    &\;\;\;\;\sum_{j\in \tilde{\scr L}_i(t)} \|x_i(t)-x_j(t)\|^2 \\
    &\ge\; ( n - 2f - (n-k) ) \mu^2 \|x_i(t)-x^*\|^2 \\
    &=\; ( k - 2f) \mu^2 \|x_i(t)-x^*\|^2.
\end{align*}
Then, 
\begin{align*}
    &\;\;\;\;\sum_{j\in\scr L_i(t)}\|x_i(t)-x_j(t)\|^2\\
    &=\sum_{j\in \tilde{\scr L}_i(t) } \|x_i(t)-x_j(t)\|^2 + \sum_{j\in \hat{\scr L}_i(t)} \|x_i(t)-x_j(t)\|^2 \\
    &\ge\;(k-2f+1)\mu^2\|x_i(t)-x^*\|^2.
\end{align*}
By applying to \eqref{eq:Si_upp_1} in Lemma~\ref{lemma:lower_bound_S}, \begin{align*}
    &\;\;\;\;\;S_i(t)\\
    &\ge\Big(\frac{(k-2f+1)\mu^2}{2}-2(|\scr H_i|-|\scr L_i(t)|)\Big)\cdot\|x_i(t)-x^*\|^2\\
    &\ge\Big(\frac{\mu^2}{2}k-\frac{4f+2f\mu^2-\mu^2}{2}\Big)\cdot\|x_i(t)-x^*\|^2.
\end{align*}
where we use $|\scr H_i|-|\scr L_i(t)| \le f$ in the last equality.
In addition, using \eqref{phi&Si}, we obtain that
\begin{align*}
    \sum_{i\in\scr H}\phi_i(t)=\sum_{i\in\scr H}S_i(t)\ge\Big(\frac{\mu^2}{2}k-\frac{4f+2f\mu^2-\mu^2}{2}\Big)V(t).
\end{align*}
This completes the proof.
\hfill $\qed$

\vspace{.05in}

\begin{lemma} \label{lemma:upper_psi}
For all time $t$,
\begin{align}\label{eq:psi_upperb}
    \sum_{i\in\scr H}\psi_i(t) &\le 4|\scr H|^3V(t).
\end{align}
\end{lemma}


{\em Proof of Lemma~\ref{lemma:upper_psi}:}
From \eqref{eq:single_sum_split} and \eqref{Bequality}, 
\begin{align*}
    &\;\;\;\;\;\Big\| \sum_{j\in\scr M_i(t)}(x_i(t)-x_{ji}(t))\Big\|\\
    &\le\sum_{j\in\scr L_i(t)}\|x_i(t)-x_j(t)\|+\sum_{j\in\scr F_i(t)}\|x_i(t)-x_{ji}(t)\|\\
    &\le \sum_{j\in\scr L_i(t)}\|x_i(t)-x_j(t)\|+\sum_{j\in\scr H_i\setminus\scr L_i(t)}\|x_i(t)-x_j(t)\|\\
    &=\sum_{j\in\scr H_i}\|x_i(t)-x_j(t)\|\\
    &\le \sum_{j\in \scr H_i}\|x_i(t)-x^*\|+\sum_{j\in\scr H_i}\|x_j(t)-x^*\|.
\end{align*}
Thus,
\begin{align*}
     &\;\;\;\;\;\sum_{i\in\scr H} \Big\| \sum_{j\in\scr M_i(t)}(x_i(t)-x_{ji}(t))\Big\|\\
     & \le (|\scr H|-1)\sum_{i\in\scr H}\|x_i(t)-x^*\|+\sum_{i\in\scr H}\sum_{j\in\scr H_i}\|x_j(t)-x^*\|.
\end{align*}
Since graph $\bbb G$ is complete, we have
\begin{align*}
     &\;\;\;\;\;\sum_{i\in\scr H}\sum_{j\in\scr H_i}\|x_j(t)-x^*\|=\sum_{i,j\in\scr H,i\neq j}\|x_j(t)-x^*\|\\
     &=\sum_{j\in\scr H}\sum_{i\in\scr H_j}\|x_j(t)-x^*\|=(|\scr H|-1)\sum_{j\in\scr H}\|x_j(t)-x^*\|.
\end{align*}
Then,
\begin{align*}
    \sum_{i\in\scr H} \Big\| \sum_{j\in\scr M_i(t)}x_i(t)-x_{ji}(t)\Big\|< 2|\scr H|\sum_{i\in\scr H}\|x_i(t)-x^*\|.
\end{align*}
Therefore, by using the definition of $\psi_i(t)$ in \eqref{eq:psi}, 
\begin{align*}
    \sum_{i\in\scr H}\psi_i(t) &=\sum_{i\in\scr H}\Big\|\sum_{j\in\scr M_i(t)}(x_i(t)-x_{ji}(t))\Big\|^2\\
    &\le \Big(\sum_{i\in\scr H}\Big\|\sum_{j\in\scr M_i(t)}(x_i(t)-x_{ji}(t))\Big\|\Big)^2\nonumber\\
    &\le 4|\scr H|^2\Big(\sum_{i\in\scr H}\|x_i(t)-x^*\|\Big)^2\\
    &\le 4|\scr H|^3\sum_{i\in\scr H}\|x_i(t)-x^*\|^2=4|\scr H|^3V(t).
\end{align*}
This completes the proof.
\hfill$\qed$

\vspace{.05in}

We are now in a position to prove Theorem~\ref{thm:single}.

\vspace{.05in}

{\em Proof of Theorem~\ref{thm:single}:}
From \eqref{eq:V:iteration}, \eqref{eq:lower_bound_sum_phi} in Lemma~\ref{lemma:single_pt_set_phi}, and \eqref{eq:psi_upperb} in Lemma~\ref{lemma:upper_psi}, we have
\begin{align*}
    &\;\;\;\;\; V(t+1)\\
    &\le V(t)-\Big(\mu^2k-4f+2f\mu^2-\mu^2\Big)\alpha V(t)+4|\scr H|^3\alpha^2V(t)\\
    &=\Big(1-\left(\mu^2k-4f-2f\mu^2+\mu^2\right)\alpha+4|\scr H|^3\alpha^2\Big)V(t).
\end{align*}
Let $\rho=1-\left(\mu^2k-4f-2f\mu^2+\mu^2\right)\alpha+4|\scr H|^3\alpha^2$. When 
$
    \alpha < (\mu^2k-4f-2f\mu^2+\mu^2)/(4|\scr H|^3),
$
$V(t)$ converges exponentially fast at a rate of $\rho$.
\hfill$\qed$




\section{Conclusion}

Resilient high-dimensional resilient constrained consensus in the presence of Byzantine agents has been studied for complete graphs in this paper. A necessary condition on feasibility redundancy for solving the problem has been provided. A sufficient condition for exponentially fast consensus has been derived for the case when the intersection of all convex constrained sets is a singleton. We observe that the derived sufficient condition on feasibility redundancy could be quite conservative compared with the necessary condition. 
For future work, we aim to reduce the gap between the necessary condition and sufficient condition, study general incomplete graphs, perform simulations for various Byzantine strategies, and apply resilient constrained consensus to distributed autonomous systems.


\bibliographystyle{unsrt}
\bibliography{ji,proposal,resilience,ref_weiguo,note,hoffman}
\newpage

\end{document}